\documentclass[11pt,twoside]{article}
\usepackage{asp2004}
\usepackage{psfig}
\usepackage{epsf}
\usepackage{graphics}
\usepackage{lscape}
\def\edcomment#1{\iffalse\marginpar{\raggedright\sl#1\/}\else\relax\fi}
\reversemarginpar

\begin{document}
\title{Supernova search at intermediate \boldmath $z$\\I. Spectroscopic analysis}  
\author{G.~Altavilla~(1), P.~Ruiz-Lapuente~(1), A.~Balastegui~(1), J.~M\'endez~(1,2), S.~Benetti (3),
M.~Irwin~(4), K.~Schahmaneche~(5), C.~Balland~(6,7), R.~Pain~(5), N.~Walton~(4)}
\affil{
1) Department of Astronomy, CER for Astrophysics, Particle
Physics and Cosmology, University of Barcelona, Diagonal 647, E--08028, 
Barcelona, Spain \\
2) Isaac Newton Group of Telescopes, 38700 Santa Cruz de La Palma, 
Islas Canarias, Spain \\
3) INAF-Astronomical Observatory of Padova, vicolo dell'Osservatorio 5, 
35122 Padova, Italy\\
4) Institute of Astronomy, University of Cambridge, Madingley Road, 
Cambridge. CB3 0HA, United Kingdom\\
5) LPNHE-IN2P3-CNRS-Universit\'es Paris 6 et Paris 7, 4 place Jussieu, 
75252 Paris Cedex  05 France\\
6) Institut d'Astrophysique Spatiale, B\^{a}timent 121, Universit\'e
Paris 11, 91405 Orsay Cedex, France\\
7) Universit\'e Paris Sud, IAS-CNAS, B\^{a}timent 121,
Orsay Cedex, France}

\begin{abstract}
We study 8 supernovae discovered  as part of the 
International Time Programme (ITP) project
``$\Omega$ and $\Lambda$ from Supernovae and the Physics
 of Supernova Explosions''
at the European Northern Observatory (ENO).
The goal of the project is to increase the sample of intermediate redshift 
($0.1<z<0.4$) SNe Ia for testing properties of SNe Ia along z and
for enlarging the sample in the Hubble diagram up to large z. 

\end{abstract}
\thispagestyle{plain}

\section{Intermediate \boldmath $z$ Supernovae }
\begin{figure}[!h]
   \plottwo{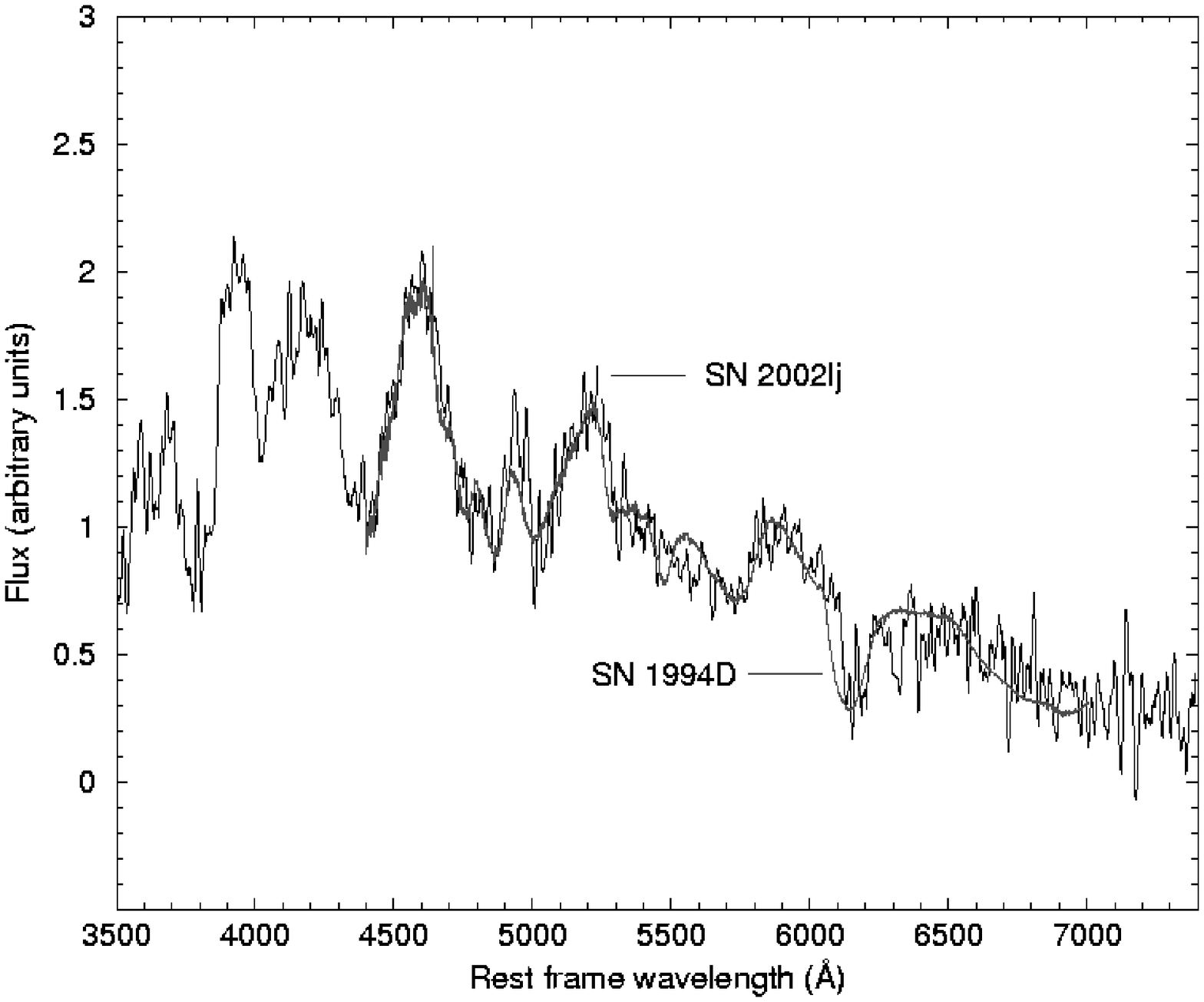}{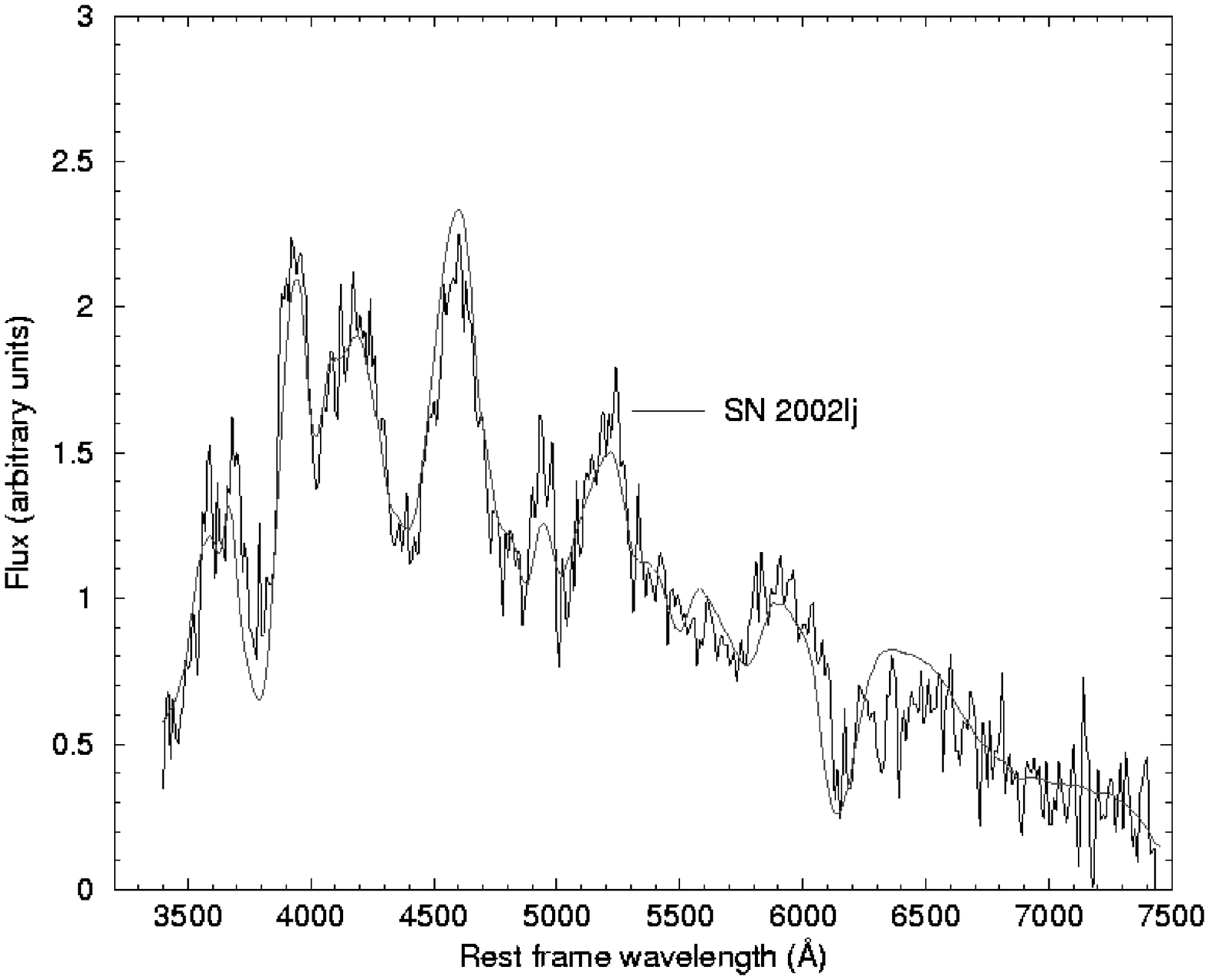}
   \caption{Comparison of SN 2002lj smoothed spectrum with that of SN 1994D 7 days past maximum 
    \citep{patat96} (left panel) and with a template  at the same epoch  \cite{serena} (right panel).}
   \label{hoddle94D} 
\end{figure}

Though SNe Ia are extensively used in cosmological studies as distance
indicators, the
empirical luminosity calibration, which is crucial for any investigation,
is based on the local SN sample. Though extensive comparison between
high-z and the local SNeIa sample is done, the intermediate--z domain (redshifts
between 0.1--0.4)  is still relatively unexplored. 
The aim of our project is to fill up the gap between the local  and the high-z
sample and to provide detailed spectroscopic and photometric data to 
improve the knowledge of SNeIa at all z.
Our observations led to the confirmation of  6 SNe Ia in the range 
 $0.033<z<0.329$, 
one of  them  being a peculiar subluminous object, and 2 SNe II, both at $z\sim 0.14$ 
(see Table \ref{tabla}). The rest  of the discovered objects were QSOs and active galaxies.
Searches were performed with the use of the 
Wide Field Camera (WFC) at the 2.5m Isaac Newton Telescope
(INT). Detection of the SN candidates was followed by
 spectroscopic identification
performed with the Intermediate dispersion Spectrograph 
and Imaging System (ISIS) mounted at the 4.2m William Herschel Telescope (WHT).
The confirmed SNe were then followed photometrically through various ENO telescopes.
   \begin{table}
     \centering
      \caption[]{Summary of observations.}
         \label{tabla}
          \footnotesize

     $$ 
         \setlength\tabcolsep{2pt}
         \begin{tabular}{l l l l l  l  l  l}
	    \hline
            \noalign{\smallskip}
 SN       & Observ. & R.A.(2000.0)                    & ~~$\delta$(2000.0)     & $z$~ & SN    & Comments \\
 name     &  date   &   ~  $^{hh}$ $^{mm}$ $^{ss}$    & ~~~~~$^{\circ}$ ' ''    &           & Type     &    \& best matching nearby SN    \\
            \hline
            \noalign{\smallskip}
2002li  & Jun. 10             & 15:59:03.08 & +54:18:16.0  ~  & 0.329   &  Ia     &  $\sim$1 week before maximum, 1990N   \\
2002lj  & Jun. 11             & 16:19:19.65 & +53:09:54.2    & 0.180    &  Ia     & $\sim$1 week past maximum, 1994D      \\
2002lp  & Jun. 10             & 16:40:11.45 & +42:28:30.2    & 0.144    &  Ia     &  around maximum, 1981B                \\
2002lq  & Jun. 10             & 16:40:28.83 & +41:14:09.1    & 0.269    &  Ia     &  $\sim$1 week before maximum, 1990N   \\
2002lr  & Jun. 10             & 22:33:12.59 & +01:05:56.7    & 0.255    &  Ia     &  $\sim$10 days past maximum, 1994D    \\
2002lk  & Jun. 10          & 16:06:55.92 & +55:28:18.2    & 0.033    &  Ia pec.& ~a few days before maximum, 1991bg            \\
 2002ln  & Jun. 10             & 16:39:24.93 & +41:47:29.0    & 0.138    & SN II     &  $\sim$2 weeks past maximum, 1999em    \\
 2002lo  & Jun. 11             & 16:39:56.42 & +42:19:20.5    & 0.136    & SN II     &  $\sim$1 month past maximum, 1999em   \\
            \noalign{\smallskip}
            \hline
         \end{tabular}
     $$ 
\end{table}
\\
The comparison of our intermediate z SN spectra with
 high-signal-to-noise ratio 
spectra of nearby SNe  has not revealed significant differences.
In particular we found a good match with nearby SN spectra
 for all our objects (see   Table \ref{tabla} and  Fig. \ref{hoddle94D}).
 The only supernova still lacking a good enough match is the
 subluminous peculiar SN 2002lk which was caught at a premaximum phase
 not commonly observed in nearby subluminous SNe Ia. 

Our overall results suggest that 
SNe Ia observed at different redshifts resemble the nearby
objects in most measured aspects (spectra, light curves, kinematics).
A further disucussion will be given elsewhere. 

\begin{thebibliography}{}
\bibitem[Nobili et al. 2003]{serena} Nobili, S., Goobar, A., Knop, R. et al. 2003, A\&A, 404, 901
\bibitem[Patat et al. 1996]{patat96} Patat, F., Benetti, S., Cappellaro, E. et al. 1996, MNRAS, 278, 111
\end{thebibliography}
\end{document}